\documentclass[aps,prd,twocolumn,showpacs,preprintnumbers,amsmath,amssymb,superscriptaddress,reprint]{revtex4-1}

\usepackage{graphicx,color} 

\newcommand{\gev}{\,\mathrm{GeV}}

\newcommand{\dmsq}{\Delta m^{2}}

\newcommand{\numu}{\nu_{\mu}}
\newcommand{\numubar}{\bar{\nu}_{\mu}}
\newcommand{\nue}{\nu_e}

\newcommand{\stwot}{\sin^2 2 \theta}

\newcommand{\evsq}{\mathrm{eV}^{2}}

\newcommand{\gtwid}{\mathrel{\raise.3ex\hbox{$>$\kern-.75em\lower1ex\hbox{$\sim$}}}}
\newcommand{\ltwid}{\mathrel{\raise.3ex\hbox{$<$\kern-.75em\lower1ex\hbox{$\sim$}}}}

\begin{document}


\title{Dual baseline search for muon neutrino disappearance at $ 0.5~\evsq < \dmsq < 40~\evsq$}

\affiliation{University of Alabama, Tuscaloosa, Alabama 35487, USA}
\affiliation{Argonne National Laboratory, Argonne, Illinois 60439, USA}
\affiliation{Institut de Fisica d'Altes Energies, Universitat Autonoma de Barcelona, E-08193 Bellaterra (Barcelona), Spain}
\affiliation{Bucknell University, Lewisburg, Pennsylvania 17837, USA}
\affiliation{University of Cincinnati, Cincinnati, Ohio 45221, USA}
\affiliation{University of Colorado, Boulder, Colorado 80309, USA}
\affiliation{Columbia University, New York, New York 10027, USA}
\affiliation{Embry Riddle Aeronautical University, Prescott, Arizona 86301, USA}
\affiliation{Fermi National Accelerator Laboratory, Batavia, Illinois 60510, USA}
\affiliation{University of Florida, Gainesville, Florida 32611, USA}
\affiliation{High Energy Accelerator Research Organization (KEK), Tsukuba, Ibaraki 305-0801, Japan}
\affiliation{Imperial College London, London SW7 2AZ, United Kingdom}
\affiliation{Indiana University, Bloomington, Indiana 47405, USA}
\affiliation{Kamioka Observatory, Institute for Cosmic Ray Research, University of Tokyo, Gifu 506-1205, Japan}
\affiliation{Research Center for Cosmic Neutrinos, Institute for Cosmic Ray Research, University of Tokyo, Kashiwa, Chiba 277-8582, Japan}
\affiliation{Kyoto University, Kyoto 606-8502, Japan}
\affiliation{Los Alamos National Laboratory, Los Alamos, New Mexico 87545, USA}
\affiliation{Louisiana State University, Baton Rouge, Louisiana 70803, USA}
\affiliation{Instituto de Ciencias Nucleares, Universidad Nacional Aut\'{o}noma de M\'{e}xico, D.F. 04510, Mexico}
\affiliation{Massachusetts Institute of Technology, Cambridge, Massachusetts 02139, USA}
\affiliation{University of Michigan, Ann Arbor, Michigan 48109, USA}
\affiliation{Princeton University, Princeton, New Jersey 08544, USA}
\affiliation{Purdue University Calumet, Hammond, Indiana 46323, USA}
\affiliation{Universit$\grave{a}$ di Roma Sapienza, Dipartimento di Fisica and INFN, I-00185 Rome, Italy}
\affiliation{Saint Mary's University of Minnesota, Winona, Minnesota 55987, USA}
\affiliation{Tokyo Institute of Technology, Tokyo 152-8551, Japan}
\affiliation{Instituto de Fisica Corpuscular, Universidad de Valencia and CSIC, E-46071 Valencia, Spain}
\affiliation{Virginia Polytechnic Institute \& State University, Blacksburg, Virginia 24061, USA}
\affiliation{Yale University, New Haven, Connecticut 06520, USA}

\author{K.~B.~M.~Mahn}
\affiliation{Columbia University, New York, New York 10027, USA}
\author{Y.~Nakajima}
\affiliation{Kyoto University, Kyoto 606-8502, Japan}
\author{A.~A. Aguilar-Arevalo}\affiliation{Instituto de Ciencias Nucleares, Universidad Nacional Aut\'{o}noma de M\'{e}xico, D.F. 04510, Mexico}
\author{J.~L.~Alcaraz-Aunion}\affiliation{Institut de Fisica d'Altes Energies, Universitat Autonoma de Barcelona, E-08193 Bellaterra (Barcelona), Spain}
\author{C.~E.~Anderson}\affiliation{Yale University, New Haven, Connecticut 06520, USA}
\author{A.~O.~Bazarko}\affiliation{Princeton University, Princeton, New Jersey 08544, USA}
\author{S.~J.~Brice}\affiliation{Fermi National Accelerator Laboratory, Batavia, Illinois 60510, USA}
\author{B.~C.~Brown}\affiliation{Fermi National Accelerator Laboratory, Batavia, Illinois 60510, USA}
\author{L.~Bugel}\affiliation{Massachusetts Institute of Technology, Cambridge, Massachusetts 02139, USA}
\author{J.~Cao}\affiliation{University of Michigan, Ann Arbor, Michigan 48109, USA}
% ...

\collaboration{MiniBooNE and SciBooNE Collaborations}\noaffiliation

\begin{abstract}
The SciBooNE and MiniBooNE collaborations report the results of a $\numu$ disappearance search in the $\dmsq$ region of $0.5-40~\evsq$. The neutrino rate as measured by the SciBooNE tracking detectors is used to constrain the rate at the MiniBooNE Cherenkov detector in the first joint analysis of data from both collaborations. 
Two separate analyses of the combined data samples set 90\% confidence level (CL) limits on $\numu$ disappearance in the $0.5-40~\evsq$ $\dmsq$ region, with an improvement over previous experimental constraints between 10 and $30~\evsq$.
\end{abstract}

\date{\today}

\preprint{FERMILAB-PUB-11-304-E}


\maketitle

\section{introduction}

Observations of neutrino oscillations at the mass splittings ($\dmsq$) of $\sim 10^{-5}~\evsq$ and $\sim 10^{-3}~\evsq$ are consistent with three generations of neutrinos and a unitary mixing matrix~\cite{Nakamura:2010zzi}. 
However, the LSND~\cite{Aguilar:2001ty} and MiniBooNE~\cite{AguilarArevalo:2010wv} experiments observe an excess of $\overline{\nu}_{e}$ events from a $\overline{\nu}_{\mu}$ beam, indicating possible new physics with $\dmsq$  $\sim 1~\evsq$. 
This could be explained by introducing additional generations of light neutrinos, and
electroweak data~\cite{:2005ema} require such additional light neutrino species to be sterile, i.e. with negligible couplings to $W^{\pm}$ and Z bosons.
 
Sensitive searches for the disappearance of $\nu_{\mu}$ (and $\overline{\nu}_{\mu}$) into sterile neutrino species can constrain models used to explain the LSND and MiniBooNE $\overline{\nu}_{e}$ appearance results. First, neutrino and antineutrino oscillations are expected to be either identical in such models or related via CP-violating phases~\cite{Karagiorgi:2006jf} or non-standard matter-like effects~\cite{Nelson:2007yq,Karagiorgi:2011ut}. Second, tests of $\nu_{\mu}$ and $\overline{\nu}_{\mu}$ disappearance ($ \stackrel{ \scriptscriptstyle ( - ) }{\numu} \rightarrow  \stackrel{ \scriptscriptstyle ( - ) }{\nu_x} $) probe elements in the neutrino mixing matrix that also govern $\nu_e$ appearance (~$\stackrel{ \scriptscriptstyle ( - ) }{\numu} \rightarrow  \stackrel{ \scriptscriptstyle ( - ) }{\nu_e}$~). As a result, a number of global analyses have been performed to study whether neutrino and antineutrino data sets relevant to the $\dmsq \sim 1~\evsq$ oscillation region in three different channels ($\nu_e$ appearance, $\numu$ disappearance, $\nu_e$ disappearance) can all be accommodated within sterile neutrino models~\cite{Karagiorgi:2009nb,Kopp:2011qd,Giunti:2011gz,Karagiorgi:2011ut}.

The MiniBooNE collaboration has previously reported limits on $\nu_{\mu}$ and  $\overline{\nu}_{\mu}$ disappearance in this region of $\dmsq$~\cite{AguilarArevalo:2009yj}.  
Substantial neutrino flux and interaction cross section uncertainties limited the sensitivity of the MiniBooNE only analyses. 
Data from the SciBooNE experiment can be used to reduce these errors because
SciBooNE shares the same neutrino flux and target material. This paper describes the $\nu_{\mu}$ disappearance analysis using data from both experiments.

\section{Experimental Apparatus}

The muon neutrino beam, with mean energy of 0.8~GeV, interacts in the SciBooNE (MiniBooNE) detectors at 100~m (541~m) from the neutrino production target. Neutrino disappearance at these distances would be observable as a distortion in the neutrino energy spectrum and as a deficit of the total event rate. At $\dmsq$ of $\sim 1\evsq$, the MiniBooNE neutrino spectrum will have a deficit relative to SciBooNE, where the neutrinos have not yet oscillated. Above $\sim 1~\evsq$, both the SciBooNE and MiniBooNE neutrino energy spectra will be altered due to disappearance.

The Fermilab Booster Neutrino Beamline (BNB) provides the $\numu$ flux to both experiments. Protons at kinetic energy of 8~GeV interact with a 1.7~interaction length beryllium target. The mesons are then focused by a magnetic field and decay in a 50~m long tunnel to produce the 93.6\% pure $\numu$ beam (5.9\% $\numubar$, 0.5\% $\nue$). The BNB neutrino flux~\cite{AguilarArevalo:2008yp} is simulated using GEANT4~\cite{Agostinelli:2002hh} and includes updated p-Be particle production 
data~\cite{:2007gt,:2007nb,*Abbott:1991en,*Allaby:1970jt,*Dekkers:1965zz,*Marmer:1969if,*Eichten:1972nw,*Aleshin:1977bb,*Sibirtsev:1988gr}.

The SciBooNE experimental hall houses three sub-detectors; in order from upstream to downstream they are a fully active and finely segmented scintillator tracker (SciBar), an electromagnetic calorimeter (EC), and a muon range detector (MRD). The SciBar detector~\cite{Nitta:2004nt} consists of 14,336 extruded polystyrene ($\rm C_8H_8$) strips. The scintillators are arranged vertically and horizontally to construct a 3 $\times$ 3 $\times$ 1.7~m$^3$ volume with a total mass of 15 tons which serves as the primary neutrino interaction target for SciBooNE. 
Each scintillator strip is read out by a 64-channel multi-anode photomultiplier tube (PMT)
via wavelength shifting fibers.
The EC is a ``spaghetti''-type calorimeter; 64 modules made of 1~mm scintillating fibers embedded in lead foil are bundled in 64 modules and read out at both ends by PMTs.
The MRD is built from 12 iron plates, each 5~cm thick, sandwiched between planes of 6~mm thick scintillation counters; there are 13 alternating horizontal and vertical planes read out via 362 individual 2~inch PMTs. A GEANT4~\cite{Agostinelli:2002hh} Monte Carlo (MC) simulation is used to model particle and light propagation through SciBooNE and the surrounding material. 

The MiniBooNE detector~\cite{AguilarArevalo:2008qa} is a mineral oil Cherenkov light detector. It is a 12~m diameter spherical tank lined with 1280 inward-facing 8~inch PMTs with an optically isolated outer region used to reject cosmic-ray induced events. Neutrino interactions in the oil produce charged particles which cause emission of primarily Cherenkov light and a smaller amount of scintillation light. A GEANT3~\cite{Brun:1987ma} based simulation, tuned on MiniBooNE and external data, is used to model light and particle production and propagation within the detector.

Neutrino interactions are simulated with the v3 NUANCE event generator~\cite{Casper:2002sd}, with cross section parameters set to match MiniBooNE and SciBooNE data~\cite{Nakajima:2010fp}.

\section{Event selection and reconstruction}

Prior to selection, approximately 44\% (39\%) of all events in SciBooNE (MiniBooNE) are due to charged current quasi-elastic (CCQE) scattering and 25\% (29\%) to charged current single pion production (CC$1\pi$); 
the small difference in event fractions in MiniBooNE and SciBooNE is due to the somewhat higher energy flux at MiniBooNE.
While all $\nu_\mu$ CC and NC interactions are sensitive to oscillation into sterile states, CCQE is the dominant interaction mode after selection cuts. The reconstructed neutrino energy ($E_{\nu}^{QE}$) is calculated assuming the neutrino interaction was CCQE, and assuming the target nucleon is at rest:
\begin{equation}
E_{\nu}^{QE} = \frac{2(M_n-E_B)E_{\mu}-(E_{B}^2-2M_nE_B + \Delta{M} + {M^2}_{\mu})}{2[(M_n-E_B)-E_{\mu} + p_{\mu}cos\theta_{\mu}]}
\end{equation}
where $\Delta{M}=M_{n}^2-M_{p}^2$, $M$ indicates the muon, proton or neutron mass with appropriate subscripts, $E_{B}$ is the nucleon binding energy  which is set to 34~MeV, $E_{\mu}$($p_{\mu}$) is the reconstructed muon energy (momentum), and $\theta_{\mu}$ is the reconstructed muon scattering angle with respect to the neutrino beam direction.  

Charged current (CC) neutrino interactions in SciBooNE are selected by identifying single muon tracks~\cite{Nakajima:2010fp}. 
The $p_\mu$ and $\theta_\mu$ of each muon track are reconstructed from the hits and energy loss in the SciBooNE sub-detectors.
The highest momentum track in the beam on-time window is required to have $p_\mu>0.25~\gev/c$ to reduce the number of neutral current (NC) events.
The energy loss of the track in SciBar must be consistent with a muon hypothesis, and must originate within the 10.6~ton SciBar fiducial volume (FV).
 Muon candidate tracks, where the endpoint of the track is also in the SciBar FV, are classified as SciBar-stopped tracks; tracks which stop in the MRD 
constitute the MRD-stopped sample. Tracks which exit the last layer of the MRD form the MRD-penetrated sample.

\begin{figure}[htbp]
\centering\includegraphics[width = \columnwidth]{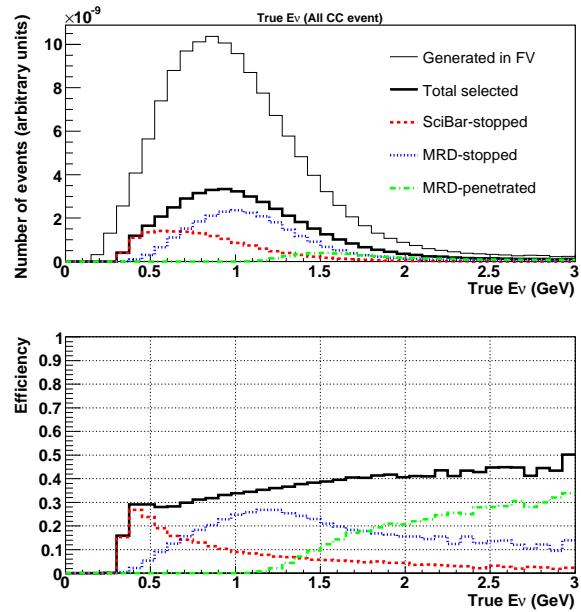}
\caption{
    (Top) Predicted number of CC events in the SciBar FV as a function of true  $E_\nu$.
    The number of selected events in each sub-sample are also shown.
    (Bottom) Detection efficiency as a function of true neutrino energy for each 
    sub-sample~\cite{Nakajima:2010fp}.}
\label{efficiency_sb}
\end{figure}

Figure~\ref{efficiency_sb} shows the event selection efficiency of each SciBooNE sample for CC interactions which occur within the SciBar FV, as estimated by MC simulation.
We estimate the selection cuts for the MRD-stopped sample are 18\% efficient for all CC interactions, with 90\% CC neutrino interaction purity; 
the main sources of inefficiency stem from the detector volume acceptance and the 
momentum requirement that the events be contained in SciBar or the MRD, or exit through the last plane of the MRD.
 The MRD-penetrated sample selection cuts are 4\% efficient and produce a 97\% pure $\numu$ CC sample.
The SciBar-stopped cuts are 12\% efficient and the sample is 85\% pure $\numu$ CC interactions.  
The mean neutrino energies of the selected events are
1.2, 2.4 and 1.0~GeV for MRD-stopped, MRD-penetrated and SciBar-stopped samples, respectively.
Together, the SciBooNE samples cover the muon kinematic region ($p_{\mu}$-$\theta_{\mu}$) relevant for events in MiniBooNE~\cite{Nakajima:2010fp}.

The selection criteria yield 13589 data events in the
SciBar-stopped sample, 20236 data events in the MRD-stopped sample, and 3544 events in the MRD-penetrated
sample after subtracting cosmic ray events, for $0.99
\times 10^{20}$ protons on target (POT) collected; with  $0.3~\gev < E_{\nu}^{QE} < 1.9~\gev$, 
there are 13592 (20166)  data events in SciBar-stopped (MRD-stopped) samples.

\begin{figure}[htbp]
\centering
\includegraphics[width = \columnwidth]{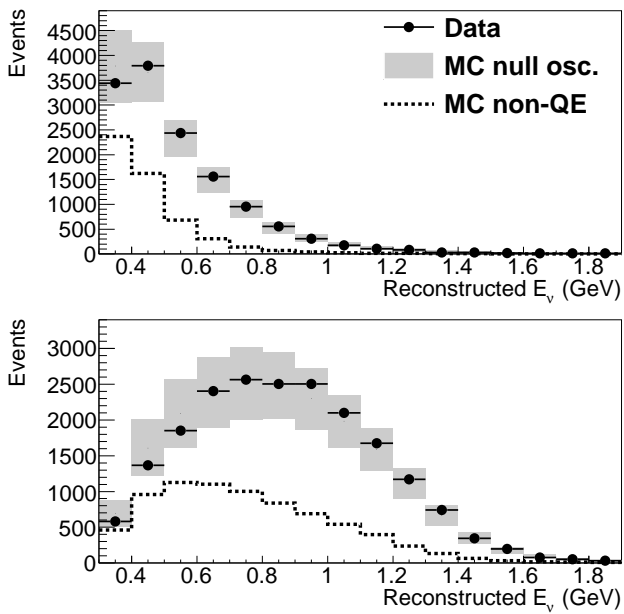}
\caption{
$E_{\nu}^{QE}$ distribution for SciBooNE data (black) with statistical errors, and prediction assuming no oscillations (relatively normalized by the total number of SciBar-stopped and MRD-stopped events) for the SciBar-stopped sample (top) and MRD-stopped sample (bottom). Cosmic background is subtracted. 
Attached to the prediction are the size of the systematic uncertainty (shaded boxes).
The predicted non-CCQE events (dash) events are also shown. 
}
\label{datamc_sb}
\end{figure}

The $E_{\nu}^{QE}$ resolution is 13\% for CCQE events in the MRD-stopped sample, and 20\% for SciBar-stopped events; the MRD-penetrated sample can only provide the muon angle for each event, not the neutrino energy.  According to the simulation, the SciBar-stopped sample is 51\% CCQE and 31\% CC$1\pi$. The remaining events are from CC or NC multi-pion events in SciBar.  The MRD-stopped sample is 52\% CCQE and 34\% CC$1\pi$; MRD-stopped sample background events also include interactions in the EC or MRD which send a muon backwards into SciBar.  The MRD-penetrated sample is 57\% CCQE and 32\% CC$1\pi$.  The small background due to cosmic events is estimated from beam-off time windows.  Fig.~\ref{datamc_sb} shows  the $E_{\nu}^{QE}$ spectrum for SciBar-stopped and MRD-stopped data and prediction, assuming no oscillations, with cosmic background subtracted. Both samples are divided into 16 bins for $0.3~\gev < E_{\nu}^{QE} < 1.9$~GeV, 0.1~GeV wide.

The MiniBooNE event selection is similar to the SciBooNE selection in that it is based on identifying a single muon in the detector~\cite{AguilarArevalo:2009yj}. Clusters of PMT hits in time are categorized as sub-events within a neutrino interaction event. The timing and charge response of the PMTs is then used to reconstruct the position, kinetic energy, and direction vector of the primary particle within each sub-event. The muon produced from CC neutrino interactions is reconstructed as the first of two sub-events, with the decay electron as the second. Exactly two sub-events are required and both sub-events must have minimal activity in the outer veto region. The first sub-event must be within the beam-on time window and have more than 200 inner tank PMT hits ($p_{\mu}> 0.25$ GeV/c) to eliminate electrons from stopped cosmic ray muon decays. The mean emission point of Cherenkov light of the first sub-event must be within the 442 ton fiducial volume. The second sub-event must have fewer than 200 inner PMT hits and the reconstructed vertex must be within 100~cm of the endpoint of the first sub-event to be consistent with an electron from the (first sub-event) muon decay.  Requiring one decay electron eliminates most NC interactions and allowing only one decay electron removes most CC charged pion interactions.

\begin{figure}[htbp]
\centering
\includegraphics[width = \columnwidth]{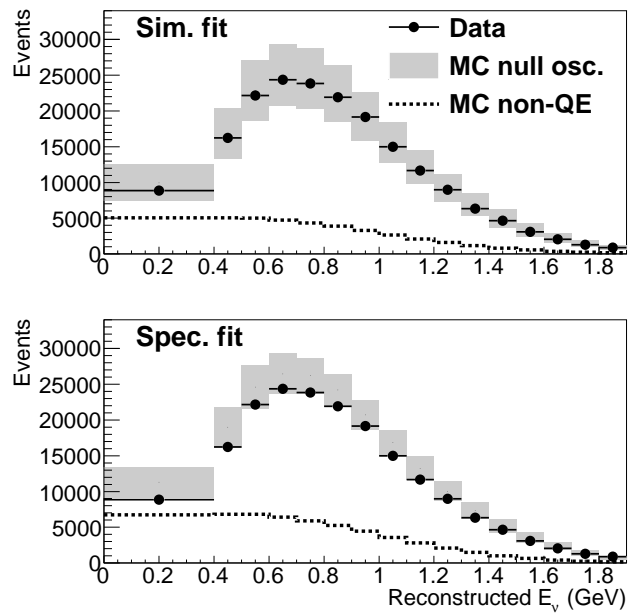}
\caption{
Top: $E_{\nu}^{QE}$ distribution for MiniBooNE data (black) with statistical errors, and prediction assuming no oscillations for the simultaneous fit analysis
(relatively normalized by the total number of SciBar-stopped and MRD-stopped events).
Attached to the prediction are the size of the systematic uncertainty (shaded boxes).
The predicted non-CCQE events (dash) events are also shown. Bottom: Same as above, for the spectrum fit analysis, where the prediction is scaled by the SciBooNE corrections.
}
\label{datamc_mb}
\end{figure}

MiniBooNE took neutrino events in two periods. The first period (Run I) comprises $5.58 \times 10^{20}$ POT, with 190,454 events with $0~\gev < E_{\nu}^{QE} < 1.9$ GeV. The second period (Run II) was taken during SciBooNE's run, comprising $0.83 \times 10^{20}$ POT and 29269 events for the same energy range. The two data sets are consistent with each other within the time dependent uncertainties, such as POT normalization, but are treated separately in the analysis, as these uncertainties cancel between MiniBooNE and SciBooNE for Run II but not for Run I.
The MiniBooNE selection is 35\% efficient with 74\% purity for CCQE events; the resultant sample has average $E_{\nu}^{QE}$ resolution of 11\% for CCQE events~\cite{AguilarArevalo:2008fb}. The rest of the $\numu$ sample is dominated by CC$1\pi$ events ($\sim$75\%) where the outgoing pion is unobserved (due, e.g., to absorption in the nucleus). Fig.~\ref{datamc_mb} shows the { the $E_{\nu}^{QE}$ spectrum} for MiniBooNE (Run I) with the prediction, assuming no oscillations, in 16 bins, 0.1 GeV wide except for the first bin (0-0.4 GeV).

\section{Search for Muon Neutrino disappearance}

\begin{table*}
\centering
  \caption{$\nu_\mu$ CC inclusive interaction rate normalization factors, with extracted uncertainties, obtained from the SciBooNE spectrum fit~\cite{Nakajima:2010fp}.}
 \label{tab:sb_spectrum}
  \begin{ruledtabular}
\begin{tabular}{ccccccc}
Energy region (GeV) & 0.25 - 0.50 & 0.50 - 0.75  & 0.75 - 1.00  &1.00 - 1.25   &1.25 - 1.75  & 1.75 -   \\ \hline
$\nu_\mu$ CC rate normalization factor ($f_i$)
& $1.65 \pm 0.22$
& $1.31 \pm 0.11$
& $1.36 \pm 0.08$
& $1.38 \pm 0.09$
& $1.36 \pm 0.12$
& $0.90 \pm 0.09$ \\
 \end{tabular}
  \end{ruledtabular}
 \end{table*}

\subsection{Analysis strategy}

Two different disappearance fits were performed on MiniBooNE and SciBooNE $\numu$ candidates.  The first method, called the simultaneous fit, is where MiniBooNE, SciBar-stopped and MRD-stopped data are fit simultaneously to test for oscillations in both detectors directly. The second method is the spectrum fit analysis, where SciBooNE data are used to extract a CC inclusive rate which is applied to MiniBooNE using scale factors as a function of true neutrino energy. The scale factors are extracted from the SciBar-stopped, MRD-stopped and MRD-penetrated samples~\cite{Nakajima:2010fp}. The corrected MiniBooNE prediction, shown in Fig.~\ref{datamc_mb}, is then compared to data with reduced uncertainties  and fitted for disappearance accounting for oscillation in both SciBooNE and MiniBooNE. 
The simultaneous fit method, by fitting both SciBooNE and MiniBooNE samples simultaneously, is advantageous by taking all correlations into account between the two experiments.  The spectrum fit, by extracting the CC interaction rate into scale factors which depend only on neutrino energy, loses some correlation between SciBooNE and MiniBooNE which reduces the power of the constraint. However, the spectrum fit demonstrates the reduction of the individual systematics at MiniBooNE due to the SciBooNE constraint, because it is done prior to the oscillation fit.
Both techniques are of interest to future experiments and so are presented here.

\subsection{Oscillation prediction}
\label{sec:pred}

For both fits, the oscillation prediction assumes two-flavor $\numu \rightarrow \nu_x$ disappearance characterized by one large mass splitting ($\dmsq \equiv \Delta m^2_{hk}$) between the active, light neutrino mass states $k$, which participate in standard three-neutrino oscillation and a fourth, heavier neutrino state $h$, and one oscillation amplitude ($\sin^22\theta = 4|U_{\mu, h}|^2(1 - |U_{\mu, h}|^2)$).

The oscillation probability, $P(\nu_{\mu} \to \nu_{x})$, is given as
 \begin{equation}
   \label{eq:oscillation_prob}
   P(\nu_{\mu} \to \nu_{x}) 
   =  \sin^{2}2\theta \sin^{2} \left( 1.27 \Delta m^{2} \frac{L}{E} \right),
 \end{equation}
where L[km] is the distance traveled and E[GeV] is the neutrino energy; $\Delta m^2$[$\mathrm{eV^{2}}$] is the mass splitting and the mixing angle, $\theta$, is dimensionless. The mean flight distance of muon neutrinos detected in SciBooNE is $\sim$76~m and in MiniBooNE is $\sim$520~m. Therefore, the oscillation probability can be non-zero for events in both SciBooNE and MiniBooNE in the oscillation search region, where neutrino energy is $\sim$1~GeV and the mass splitting ($\Delta m^2$) is  $\sim$1~eV${}^2$. The predicted number of events in each reconstructed $E_{\nu}^{QE}$ bin in each sample, $p_i(\sin^22\theta, \Delta m^2)$, is calculated assuming all $\numu$ and $\numubar$ events, both CC and NC, can oscillate according to a particular $\dmsq$-$\stwot$ point and the true neutrino energy. The incident neutrino's energy and distance from the point of its parent decay to detection are used from the simulation to properly represent the range of E$_{\nu}$, L of the SciBooNE and MiniBooNE samples.

The simultaneous fit prediction, $p_i^{sim}$, has 48 bins corresponding to the SciBooNE and MiniBooNE reconstructed $E_{\nu}^{QE}$ bins (16 SciBar-stopped bins, 16 MRD-stopped bins, and 16 MiniBooNE bins corresponding to Run I or Run II only). The prediction in each bin is normalized by the sum of the observed number of SciBar-stopped and MRD-stopped events, $d_k$, to prediction, $n_k$, assuming oscillation at $\dmsq$, $\stwot$:

\noindent
 \begin{align}
   \label{eq:simfitpred}
 &  p_i^{sim}(\sin^22\theta, \Delta m^2) \nonumber \\ 
& = \frac{\sum_{k=1}^{32} d_k}{\sum_{k=1}^{32} n_k (\sin^22\theta, \Delta m^2)} \cdot
   n_{i}(\sin^22\theta, \Delta m^2)
 \end{align}
where $k$ and $i$ are indices over reconstructed $E_{\nu}^{QE}$ bins.

For the spectrum fit, the prediction $p_i^{spec}$ corresponds to 32 MiniBooNE reconstructed $E_{\nu}^{QE}$ bins (16 Run I bins and 16 Run II bins).  The prediction in each bin is built by additionally applying the extracted rate normalization factors, $f_i$, obtained by the analysis of $\nu_\mu$ charged current interactions at SciBooNE~\cite{Nakajima:2010fp}. 
The normalization factors, $f_i$, given in Table~\ref{tab:sb_spectrum} as a function of incident neutrino energy, will be referred to as the SciBooNE corrections. As with the simultaneous fit, the prediction is further scaled by the expected number of oscillated events at SciBooNE: 

\noindent
 \begin{align}
   \label{eq:miniboone_enuqe_osc}
&  p_i^{spec}(\sin^22\theta, \Delta m^2) \nonumber \\ 
& = \sum_j^{6} f_j
   \frac{\mathcal{N}_j(0,0)}{\mathcal{N}_j(\sin^22\theta, \Delta m^2)} \cdot
  \mathcal{M}_{ij}(\sin^22\theta, \Delta m^2),
 \end{align}
where $\mathcal{N}_{j}$ is the predicted number of events in true $E_{\nu}$ bin $j$ at SciBooNE, and  $\mathcal{M}_{ij}$ is the predicted number of events at MiniBooNE for reconstructed $E_{\nu}^{QE}$ bin $i$ and incident true neutrino energy bin $j$, assuming oscillation with parameters $\dmsq$,~$\stwot$.


\subsection{Systematic Uncertainties}

\begin{table*}
\centering
 \caption{Ratio of observed number of events to predicted for each sample with uncertainties due to statistics, neutrino flux, cross section and detector systematics.
}
 \label{tab:ratio}
  \begin{ruledtabular}
\begin{tabular}{ccccccc}
                  &  & \multicolumn{5}{c}{Uncertainties (\%)}\\ \cline{3-7}
Sample           & Data/MC & Stat & Flux & Cross section & Detector & Total sys.\\ \hline
SciBar-stopped  & 1.25 & 1.0  & 6.2 & 15.3 & 2.2  & 16.6 \\
MRD-stopped     & 1.26 &  0.8 & 8.0 & 16.8 & 2.8  & 18.3 \\
MRD-penetrated  & 1.04 & 1.7 & 18.2 & 17.1 & 5.8 & 25.7 \\
MiniBooNE &  1.21  & 0.3 & 6.2 & 15.1 & 4.8 & 17.1 \\
 \end{tabular}
  \end{ruledtabular}
 \end{table*}

\begin{table*}
\centering
 \caption{Ratio of observed number of events to predicted for MiniBooNE sample with uncertainties due to statistics, neutrino flux, cross section and detector systematics, after applying the SciBooNE corrections listed in Tab.~\ref{tab:sb_spectrum}. The effect of the uncertainties associated with the extraction of the SciBooNE corrections ($f_i$) are also shown.
}
 \label{tab:ratio_afterfit}
  \begin{ruledtabular}
\begin{tabular}{cccccccc}
                  &  & \multicolumn{6}{c}{Uncertainties (\%)}\\ \cline{3-8}
Sample           & Data/MC & Stat & Flux & Cross section & Detector & SB spectrum & Total sys.\\ \hline
MiniBooNE &  0.90  & 0.2 & 0.8 & 4.7 & 4.8 & 1.7 & 7.3\\
 \end{tabular}
  \end{ruledtabular}
 \end{table*} 

\begin{table}
\centering
 \caption{Ratio of observed number of events to predicted for SciBooNE samples after re-applying the SciBooNE corrections listed in Tab.~\ref{tab:sb_spectrum}. 
}
 \label{tab:ratio_afterfit_sciboone}
  \begin{ruledtabular}
\begin{tabular}{cccc}
Sample           & SciBar-stopped & MRD-stopped & MRD-penetrated\\ \hline
Data/MC           & 0.94 & 0.97 & 0.98 \\
 \end{tabular}
  \end{ruledtabular}
 \end{table}

Systematic uncertainties are included for the underlying neutrino flux prediction, neutrino interaction cross section, and detector response. The method used to estimate the uncertainties due to the underlying neutrino flux prediction and the MiniBooNE detector model is identical to the method used in previous MiniBooNE results~\cite{AguilarArevalo:2007it,AguilarArevalo:2009yj}; the cross section uncertainties and SciBooNE detector uncertainties assumed are identical to previous SciBooNE results~\cite{Nakajima:2010fp}. 
The dominant uncertainties on the neutrino event rate at SciBooNE and MiniBooNE are the uncertainty of the neutrino flux due to production of $\pi^+$ mesons from the Be target, and the CCQE and CC1$\pi$ neutrino cross sections. The detector uncertainties affect the muon energy and angular resolution. For SciBooNE, these include uncertainties on the energy loss of muons through scintillator and iron, light attenuation in the WLS fibers and PMT response; for MiniBooNE, detector uncertainties include particle and light propagation through the mineral oil, scattering, detection and PMT response. Table~\ref{tab:ratio} shows the ratio of predicted to observed number of events for the SciBooNE and MiniBooNE samples, and the size of the flux, cross section and detector uncertainties.

\begin{figure}[htbp]
\centering
\includegraphics[height=80mm]{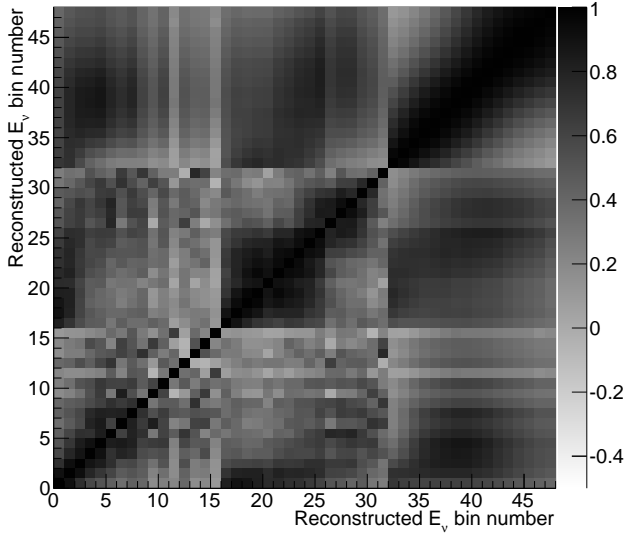}
\caption{
Correlation coefficients of the total systematic uncertainties on the reconstructed $E_\nu$
distribution for the simultaneous fit analysis.
Bins 0-15 and bins 16-31 are respectively for the SciBar-stopped 
and the MRD-stopped samples from SciBooNE, 
shown in Fig.~\ref{datamc_sb}.
Bins 32-47 are for the MiniBooNE sample shown in the top panel of Fig.~\ref{datamc_mb}.
}
\label{ematrix_simfit}
\end{figure}

\begin{figure*}[htbp]
\centering
\includegraphics[width = 2  \columnwidth]{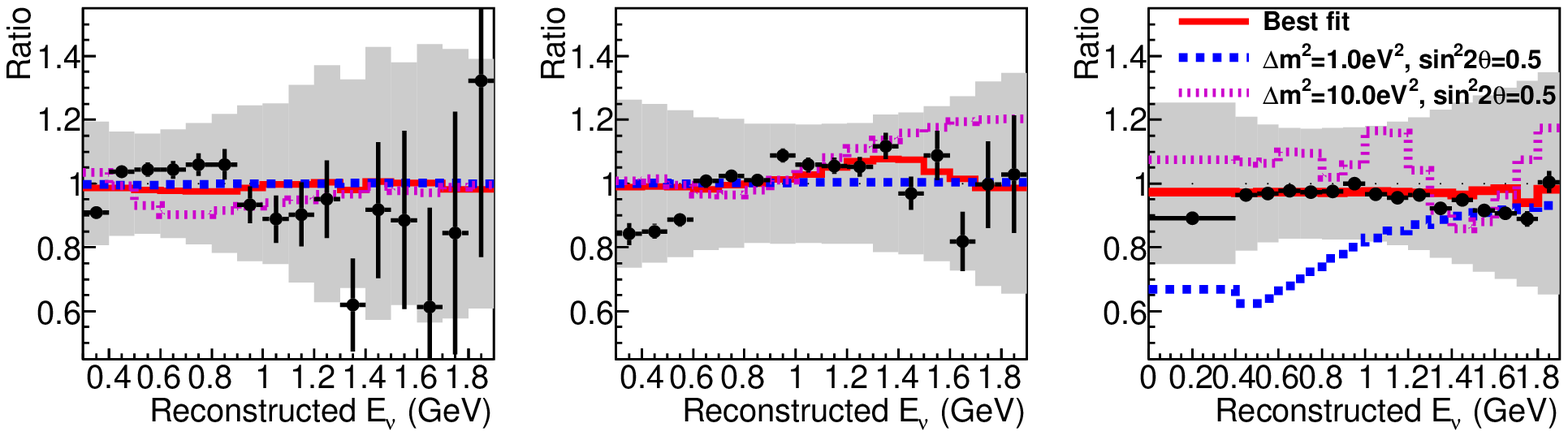}
\caption{
Ratios with respect to the null hypothesis for SciBar-stopped (left), MRD-stopped (middle) and MiniBooNE (right) samples
for the simultaneous fit analysis.
Data (points), and disappearance expectations for 
the best fit (solid red) and parameters 
$\dmsq=1.0 ~\evsq$, $\stwot=0.5$ (dashed blue) and $\dmsq=10.0 ~\evsq$, $\stwot=0.5$ (dotted magenta) are shown. 
The best fit parameters are $\dmsq=43.7 ~\evsq$, $\stwot=0.60$.
Fractional size of the diagonal elements of the error matrix for the null oscillation is shown as the gray bands.
}
\label{simfit_ratio}
\end{figure*}

\begin{figure}[htbp]
\centering
\includegraphics[width =  0.9\columnwidth]{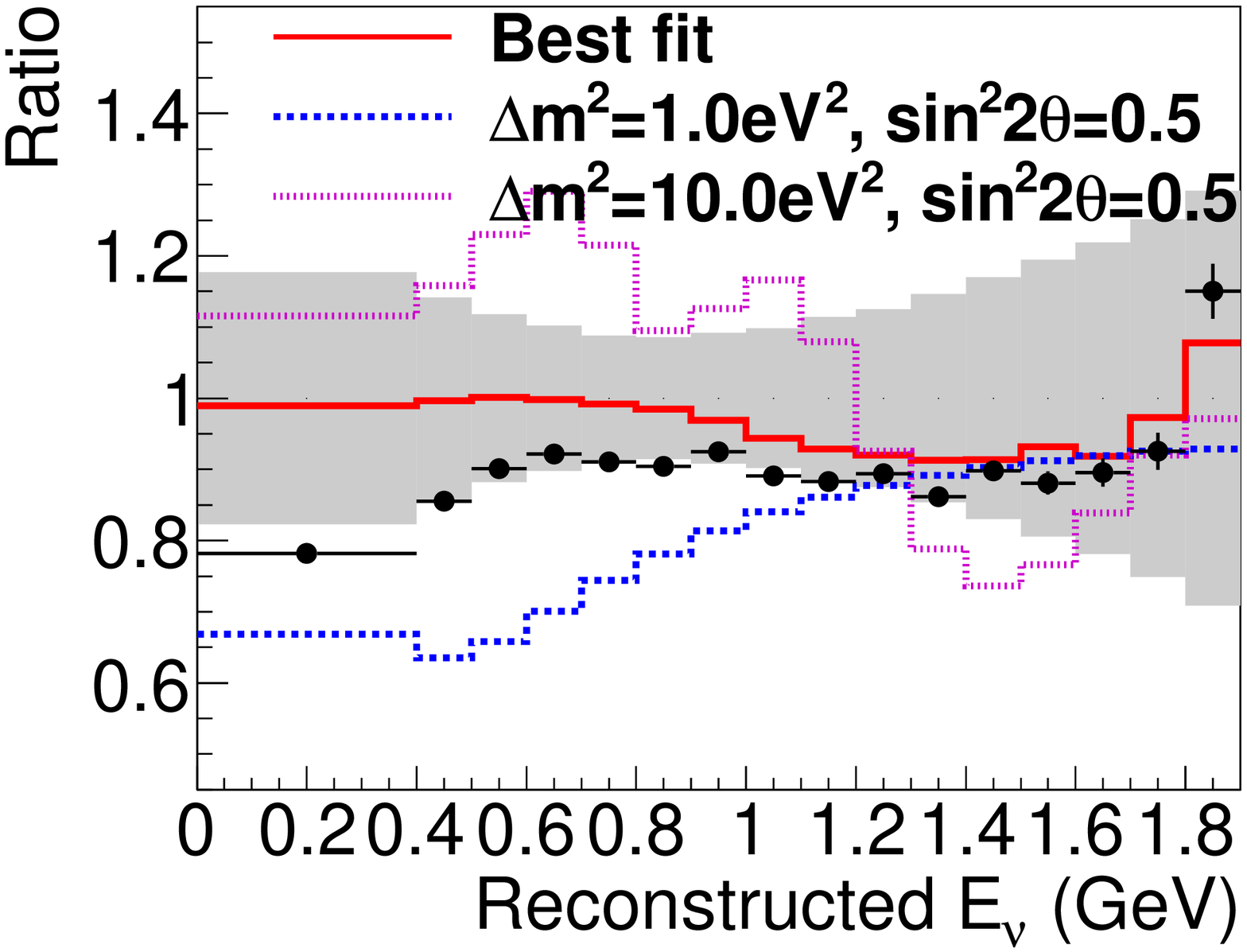}
\caption{
Ratios with respect to the null hypothesis for MiniBooNE samples for the spectrum fit analysis.
Data (points), and disappearance expectations for 
the best fit (solid red) and parameters 
$\dmsq=1.0 ~\evsq$, $\stwot=0.5$ (dashed blue) and $\dmsq=10.0 ~\evsq$, $\stwot=0.5$ (dotted magenta) are shown. 
The expectations are for the spectrum fit analysis.
The best fit parameters are $\dmsq=41.7 ~\evsq$, $\stwot=0.51$.
Fractional size of the diagonal elements of the error matrix for the null oscillation is shown as the gray bands.
}
\label{specfit_ratio}
\end{figure}

Each source of systematic uncertainty produces correlations between $E_{\nu}^{QE}$ bins within a single data sample and between samples, e.g. between SciBooNE and MiniBooNE. These correlations are represented by developing a covariance matrix, $V_{ij}$, which is the sum of the individual covariance matrices for each source of systematic uncertainty. Each covariance matrix is constructed in the same manner as in 
previous SciBooNE CC inclusive analysis~\cite{Nakajima:2010fp} and MiniBooNE oscillation analyses~\cite{AguilarArevalo:2007it,AguilarArevalo:2008rc}; 
a fractional error matrix is generated using the MC expectations for the central value ($p_{i(j)}$) 
and many different sets for systematic variations ($p_{i(j),k}^{sys}$) for each $E_\nu^{QE}$ bin: 
\begin{equation}
V_{ij} = \frac{1}{M}\sum_k^M\frac{p_{i,k}^{sys} - p_i}{p_i}\frac{p_{j,k}^{sys} - p_{j}}{p_j},
\end{equation}
where $k$ denotes the index for each systematic variation and $M$ denotes the total number 
of variations.
For the simultaneous fit the covariance matrix ranges over $E_{\nu}^{QE}$ bins from SciBar-stopped, MRD-stopped and MiniBooNE samples. The sizes of the diagonal elements of the covariance matrix, $\sigma_i=p_i\sqrt{V_{ii}}$ are shown as the error bands on the predictions for the individual samples in Figs.~\ref{datamc_sb}~and~\ref{datamc_mb}, and also in Figs.~\ref{simfit_ratio}~and~\ref{specfit_ratio}. The correlation matrix component of the covariance matrix, $V_{ij}/\sqrt{V_{ii}V_{jj}}$, is shown in Fig.~\ref{ematrix_simfit}. Where there is a shared source of uncertainty (e.g., neutrino flux or cross section), MiniBooNE bins are strongly correlated with SciBooNE bins. 

The correlations between SciBooNE and MiniBooNE are exploited by the simultaneous fit $\chi^2$, described in Sec.~\ref{sec:fit}, to constrain the rate at MiniBooNE. When a bin at MiniBooNE is strongly correlated to one at SciBooNE, a difference between the data and prediction at MiniBooNE will only increase the $\chi^2$ if the difference is not also present in the appropriate SciBooNE bin.  The differences in the event rates in the two detectors due to flux or cross section systematics, unlike detector systematics, tend to cancel.  As a result, the largest uncertainty for the simultaneous fit analysis is the MiniBooNE detector uncertainty.

\begin{figure}[htbp]
\centering
\includegraphics[height=80mm]{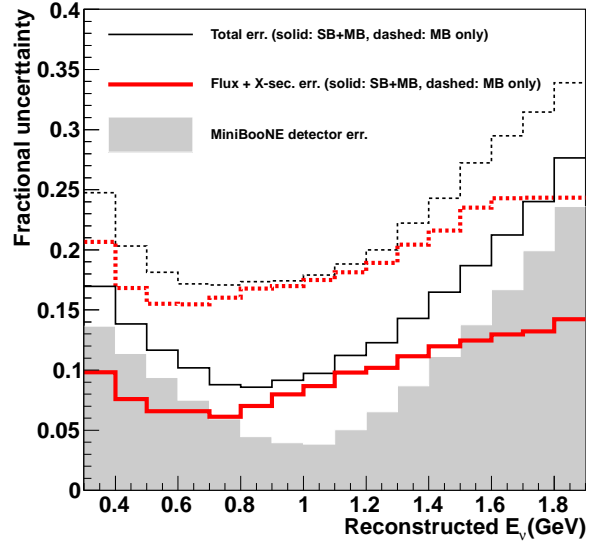}
\caption{
The fractional size of the systematic uncertainties for each MiniBooNE reconstructed $E_\nu$ bins 
for the spectrum fit analysis.  The dashed lines show the uncertainties using MiniBooNE alone, 
and the solid lines show the uncertainties after the SciBooNE corrections.
}
\label{errors}
\end{figure}

\begin{figure}[htbp]
\centering
\includegraphics[height=80mm]{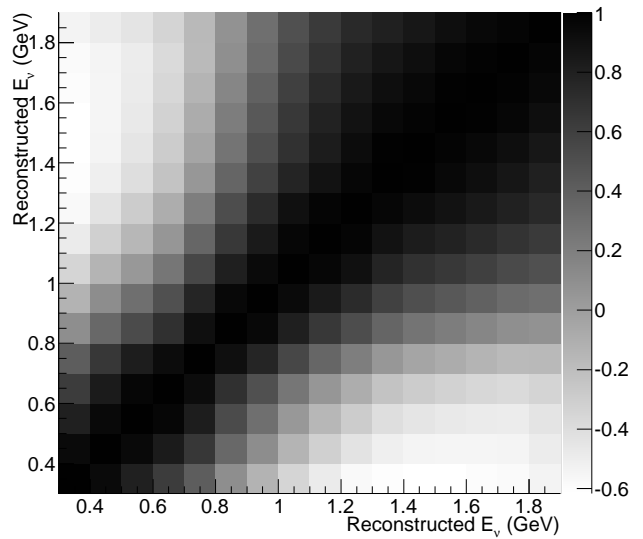}
\caption{
Correlation coefficients of the total systematic uncertainties on the MiniBooNE 
reconstructed $E_\nu$
distribution for the spectrum fit analysis, shown in the right panel 
of Fig.~\ref{datamc_mb}.
}
\label{ematrix_specfit}
\end{figure}

In the spectrum fit method, uncertainties of neutrino flux and cross sections are constrained by the SciBooNE corrections. The systematic variations of the predicted number of events in MiniBooNE after the SciBooNE corrections, $p_i^{spec,sys}$, are calculated as:
 \begin{equation}
   \label{eq:specfit_systematic}
   p_i^{spec,sys} 
 = \sum_j^{6} f_j
   \frac{\mathcal{N}_j}{\mathcal{N}_j^{sys}} \cdot
  \mathcal{M}_{ij}^{sys},
 \end{equation}
where $\mathcal{N}_j$ and $\mathcal{N}_j^{sys}$ are the central value and the systematic variation (sys), respectively, 
of the predicted number of events in SciBooNE, and  $\mathcal{M}_{ij}^{sys}$ is the systematic variation of the predicted number of events at MiniBooNE for $E_{\nu}^{QE}$ bin $i$ and incident neutrino energy bin $j$.
Then, $p_i^{spec,sys}$ is used to construct the covariance matrix for the spectrum fit analysis for MiniBooNE $E_{\nu}^{QE}$ bins.
The ratio of the observed number of events predicted for the MiniBooNE sample and the size of the reduced systematic uncertainties after applying the SciBooNE corrections are shown in Tab.~\ref{tab:ratio_afterfit}.
As a closure test, the SciBooNE corrections are re-applied to the SciBooNE MC samples and compared with the data, as shown in Tab.~\ref{tab:ratio_afterfit_sciboone}.
The data to MC ratios for the SciBooNE samples are not exactly zero due to the small data and MC disagreement at the low $Q^2$ region~\cite{Nakajima:2010fp}, however 
the differences are well within the systematic uncertainties.
The fractional sizes of the diagonal elements of the covariance matrix are shown in Fig.~\ref{errors}, and the correlation coefficients for the total systematic uncertainty after the SciBooNE corrections are shown in Fig.~\ref{ematrix_specfit} for MiniBooNE  $E_{\nu}^{QE}$ bins. After the SciBooNE corrections, the flux and cross section uncertainties are reduced by a factor of approximately two as compared to the previous MiniBooNE-only analysis. The spectrum fit method has uncertainties associated with the extraction of the $f_i$ rate factors, given in Table~\ref{tab:sb_spectrum}, which are included in $V_{ij}$. As shown in Tab.~\ref{tab:ratio_afterfit}, the dominant uncertainty for the spectrum fit analysis is the MiniBooNE detector uncertainty which does not cancel in the ratio between the two experiments.

\subsection{Oscillation Fit}
\label{sec:fit}

The disappearance search uses a frequentist-based $\Delta \chi^2$ approach~\cite{Feldman:1997qc}
 to determine allowed regions in the $\dmsq-\stwot$ plane. The $\chi^2$ is calculated from a comparison of the data, $d_i$, in $E_{\nu}^{QE}$ bin $i$, to a prediction $p_i(\dmsq,\stwot)$: 
\begin{equation}
\chi^2 = \sum_{i,j}^{E_\nu^{QE} \mathrm{bins}}\frac{d_i-p_i}{p_i} {V_{ij}}^{-1}\frac{d_j-p_j}{p_j}
\end{equation}
where $V_{ij}$ is the covariance matrix.  As defined in Section~\ref{sec:pred}, the simultaneous fit prediction $p_i^{sim}$ has 48 bins and the spectrum fit prediction, $p_i^{spec}$ has 32 bins. The covariance matrix $V_{ij}$ is either 48 $\times$ 48 bins (simultaneous fit) or 32 $\times$ 32 bins (spectrum fit).
At each oscillation point in a grid across the region of interest, $\Delta \chi^2(\dmsq,\stwot) = \chi^2(\dmsq,\stwot) - \chi^2(min)$  is calculated. The probability distribution for the $\Delta \chi^2$ statistic is determined from an analysis of a set of simulated data samples generated with neutrino flux, cross section, and detector systematics varied.

\begin{figure}[htbp]
\centering
\includegraphics[height=80mm]{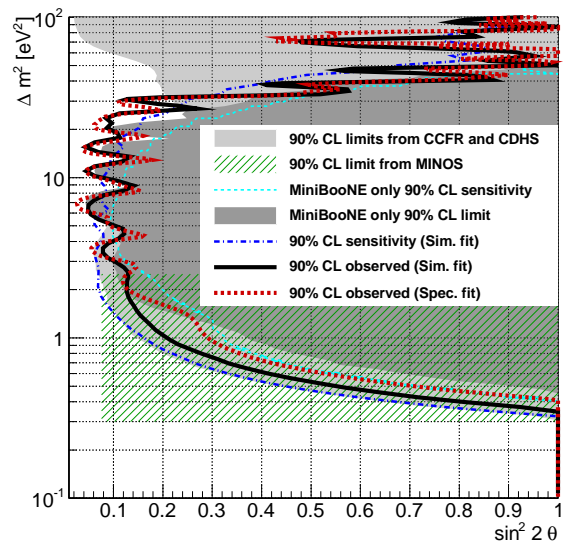}
\caption{
The 90\% CL limit for the simultaneous fit (solid black curve),
 and for the spectrum fit (red dashed curve). 
The sensitivity for the simultaneous fit analysis is also shown (dot-dash curve) with previous limits by 
CCFR~\cite{Stockdale:1984cg} and CDHS~\cite{Dydak:1983zq} (light gray), MiniBooNE~\cite{AguilarArevalo:2009yj} (dark gray) and MINOS~\cite{Adamson:2011ku} (green hash box). 
The sensitivity for the MiniBooNE-only analysis~\cite{AguilarArevalo:2009yj} 
is also shown (cyan dashed curve).
}
\label{limit}
\end{figure}

\section{Results and Conclusion}

Both the simultaneous fit and spectrum fit are consistent with the null hypothesis. Fig.~\ref{simfit_ratio} and Fig.~\ref{specfit_ratio} show the ratio of data to the null hypothesis and three oscillation scenarios for illustration, for the simultaneous fit and the spectrum fit, respectively. The $\chi^2$ value for the null hypothesis is 45.1 (48 degrees of freedom (DOF), 59\% probability), with $\chi^2(min)=39.5$ at $\dmsq=43.7 \evsq$, $\stwot=0.60$ in the simultaneous fit analysis, fit to MiniBooNE Run I data. The expected $\Delta \chi^2$ for the 90\% CL for the simultaneous fit is 9.34, so the result is consistent with the null hypothesis. A fit to Run II data is consistent with a $\chi^2(null)$ = 41.5 (48 DOF, 73\% probability); a fit to both Run I and Run II produces a negligible improvement on the simultaneous fit sensitivity as compared to fitting just the larger Run I data set. For the spectrum fit to Run I and Run II, the $\chi^2$ found comparing the data to the null hypothesis is 41.5 (32 DOF, 12\% probability) with the $\chi^2$  minimum value of 35.6 at $\dmsq=41.7 \evsq$, $\stwot=0.51$. The expected $\Delta \chi^2$ for the 90\% CL is 8.41, so the fit is consistent with the null hypothesis.

Figure~\ref{limit} shows the 90\% CL sensitivity for the simultaneous fit analysis compared to the previous MiniBooNE-only sensitivity. The 90\% CL sensitivity is the average limit of 1000 fake experiments with no underlying oscillation assumed. Both sensitivities use the MiniBooNE Run I data to show how the addition of SciBooNE data improves the sensitivity.  In the MiniBooNE-only fit, flux and cross section uncertainties are dominant at low $\dmsq$ ($\dmsq<1\evsq$), resulting in a reduced sensitivity in that region. In the simultaneous fit, the presence of flux and cross sections uncertainties does not increase the $\chi^2$ once SciBooNE data are included and the sensitivity improves. In the spectrum fit, which has a similar sensitivity as the simultaneous fit, the uncertainties are directly reduced by the SciBooNE corrections and so the sensitivity is improved.


The 90\% CL limit for both analyses is also shown in Fig.~\ref{limit} along with other relevant experimental results. While the limit is an improvement in the $\dmsq$ region of $10-30~\evsq$, the 90\% CL limits are worse than the sensitivity at $\dmsq \sim 1~\evsq$ because of the small deficit at MiniBooNE relative to SciBooNE. The energy dependence of this deficit is within the uncertainties, thus the limit is consistent with the region covered by 68\% of the limits of fake null experiments.  The spectrum fit limit is not as strong as the simultaneous fit limit because the corrected spectrum creates a larger deficit between MiniBooNE data and prediction than the default prediction, while the simultaneous fit applies no correction and so is consistent with the small deficit.

In summary, we search for $\numu$ disappearance using both SciBooNE and MiniBooNE data sets. The two analyses presented reduce the flux and cross section uncertainties to the level of the detector uncertainties of the two experiments. We set 90\% CL limits on $\numu$ disappearance in the $\dmsq$ region of $0.5-40~\evsq$, with an improvement over previous experimental results between $10$ and $30~\evsq$.

\begin{acknowledgments}
We acknowledge the support of Fermilab. We acknowledge the support of grants, contracts and fellowships from MEXT, JSPS (Japan), the INFN (Italy), the Ministry of Science and Innovation and CSIC (Spain), the STFC (UK), and the DOE and NSF (USA). 
We also acknowledge the use of CONDOR software for the analysis of the data.
\end{acknowledgments}

\end{document}